# BLIND SOURCE SEPARATION METHODS FOR DECONVOLUTION OF COMPLEX SIGNALS IN CANCER BIOLOGY

Zinovyev A.[1,2,3*,†], Kairov U.[4,5,†], Karpenyuk T.[4] and Ramanculov E.[5]

[1]Institute Curie, Paris, France
[2]INSERM U900, Paris, France
[3]Mines ParisTech, Fontainebleau, France
[4]Kazakh National University after Al-Farabi, Almaty, Kazakhstan
[5]National Center for Biotechnology of the Republic of Kazakhstan, Astana, Kazakhstan

*Corresponding author
e-mail: andrei.zinovyev@curie.fr

[†]These two authors contributed equally to the work

## Abstract

Two blind source separation methods (Independent Component Analysis and Non-negative Matrix Factorization), developed initially for signal processing in engineering, found recently a number of applications in analysis of large-scale data in molecular biology. In this short review, we present the common idea behind these methods, describe ways of implementing and applying them and point out to the advantages compared to more traditional statistical approaches. We focus more specifically on the analysis of gene expression in cancer. The review is finalized by listing available software implementations for the methods described.

## Introduction

*Linear approximation of large datasets*

In many fields of science, data can be represented as a rectangular matrix with some objects (for example, $n$ genes) corresponding to the matrix rows and the objects' features (for example, gene expression in $m$ tumor biopsies) corresponding to the matrix columns. These matrices can be huge: thus, methods for revealing patterns in the distribution of the matrix element values are of extreme use. One particular approach is connected to the idea of approximating a rectangular matrix by another matrice, having much lower rank: $X \approx AS$, where $X$ is a matrix of data of size $m \times n$, and $A$ is a $m \times k$ matrix, $k << m$. We will call the rows of the $A$ matrix *components* ($m$-dimensional vectors), and the columns of the $S$ matrix *projections* of data vectors onto the components (a $k$-dimensional vector for each of $n$ data points).

A simple optimization problem $\| X - AS \|^2 \rightarrow$ min with Euclidean metrics leads to the well-known Singular Value Decomposition (SVD) or, equivalently, Principal Component Analysis (PCA), two fundamental methods introduced in the data analysis more than one hundred years ago [1]. These methods work best in the case of multidimensional Gaussian data distribution. Several families of statistical methods were suggested in 1980-1990s implementing additional constrains put on the choice if components, including blind source separation techniques, which we will describe below (see Figure 1).

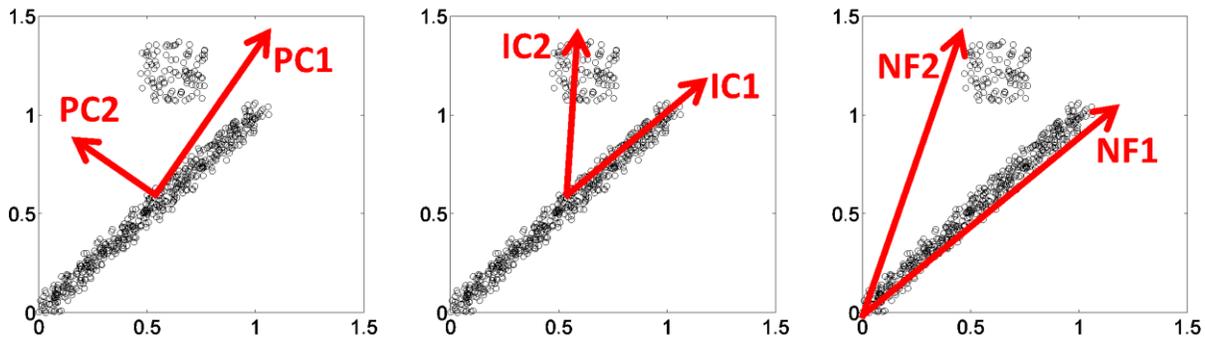

**Figure 1**. Simple example of 2D data cloud with three methods of matrix approximation applied. Here small circles are data points, PCs are principal components (PC1 explains maximum of variance, PC1 and PC2 are orthogonal), ICs are independent components (give maximally non-gaussian distribution of the projections), NFs are non-negative matrix factorization components (the data points have only positive projections on them).

*Cocktail party problem or blind source signal separation*

Cocktail party problem was initially formulated in signal processing [2]. It consists in decomposing a complex mixture of audio signals coming from different sources (at a real party it can be music, talking, singing, applauding, etc.) and mixed by several sensors (party microphones distributed around the party hall) with weights dicreasing with the distance from the sensor to the source of the signal. The problem is solvable given statistical independence of the signal sources (it is much more difficult of impossible to distinguish separate voices in a chore).

*Independent Component Analysis (ICA)*

The idea of ICA was first formulated and named by Herault and Jutten back in 1982. It was suggested using higher-order moments for matrix approximation, considering all Gaussian signals as noise. It was followed by other similar works and was rigorously and theoretically described in [3]. Efficient and fast algorithms were developed for ICA [4]. It was shown that the requirement of statistical independence is equivalent to maximizing non-gaussivity of data point projections onto the components, measured by some combination of higher data distribution moments (kurtosis) or other functions (negentropy, tangent function, etc.) The Gaussian signal contained in the data is usually subtracted from the data before application of ICA by data whitening such that all second moments become equal unity. Therefore, independent components can be non-orthogonal in the original space of data, which can be considered as an advantage in applications (Figure 1).

*Non-negative Matrix Factorization (NMF)*

The idea of solving $\| X - AS \|^2 \rightarrow \min$ problem (not necessarily with Euclidean metrics) with constraints on the sign of the *A* and *S* matrice elements was first suggested in [5] but become well-known after demonstration of its application to face recognition and word frequency analysis [6]. The requirement on the *A* and *S* elements to be positive creates the components' sparsity effect (create vectors with many close to zeros entries), unlike the unconstrained PCA analysis. This is advantageous in many applications where a decomposition of a complex mixture of signals into a sum of components is needed, with positive weight of each of the component (subtraction is forbidden).

*Advantages of linear factorization methods compared to hierarchical clustering*

Hierarchical clustering is a method of choice in the majority of analyses of high-thrughput molecular biology data [7]. Its widespread use can be explained by the simplicity of understanding the method's principle, ease of visual interpretation and availability of user-friendly software for applying it. However, in many applications, hierarchical clustering is not the most suitable method because of its known instability to a random removal of samples or features and separation of samples or features into non-intersecting groups.

Functional studies in molecular biology show that the majority of genes have pleiotropic function. Almost every gene can respond to a variety of distinct external signals. Linear factorizaion methods have an intrinsic possibility to associate a gene (sensor) to several sources of signal (biological functions) which makes it a suitable tool for analysis of complex biological data. Moreover, since the linear factorization methods are based on some kind of averaging of the data (calculating data moments), they are intrinsically more stable to the presence of high levels of noise in the data and partial removal of samples, if compared to the agglomerative clustering methods.

**Linear matrix factorization for transcriptomic data analysis in cancer biology**

*Space of genes and space of samples*

Cocktail party problem has a very close interpretation in molecular biology if the following interpretation is accepted: a gene is an analogue of a sensor which receives regulatory signals from several sources (biological factors), see Figure 2. The biological factors can be activities of transcription factors or other various influences coming from a particular intercellular context or from environment. The combination of factor activities regulates gene expression through a complex (and unknown) function. As the first approximation, we can assume that this function is linear:

$$\text{Expression}(\text{gene } i, \text{sample } s) = \Sigma_{i=1..m} a_{Fj}^{\text{gene } i} \text{Activity}_{Fj}(\text{sample } s).$$

An alternative, less mechanistic and more statistical, interpretation is also possible: hidden factors (such as patient age, sample quality, severity of diagnosis, etc.) are mixed together in a linear fashion to determine a position of a sample (for example, a tumor biopsy) in the multidimensional space with axes corresponding to the individual gene expressions.

In the first case, a point in the data space is a gene and we will call it the "gene" space. In the second case, a point is a sample and we call it the "sample" space. It is important to distinguish these two interpretations in practical applications since it corresponds to two different problem statements. Technically, this corresponds to application of blind source separation methods to the initial or transposed matrix of gene expression.

First report on application of ICA to deconvolution of signals in analysis of gene expression changes during yeast sporulation appeared in [8]. In this work, ICA components, called "modes of gene expression", reproduced manually defined biological classes of yeast genes. First applications of NMF to yeast gene expression data was reported in [9] and to cancer in [10]. Since then the number of applications of matrix factorization methods in analysis of gene expression has grown very rapidly. There exist several reviews, comparing different methods and pointing out at their advantages and disadvantages [11, 1, 12, 13, 14].

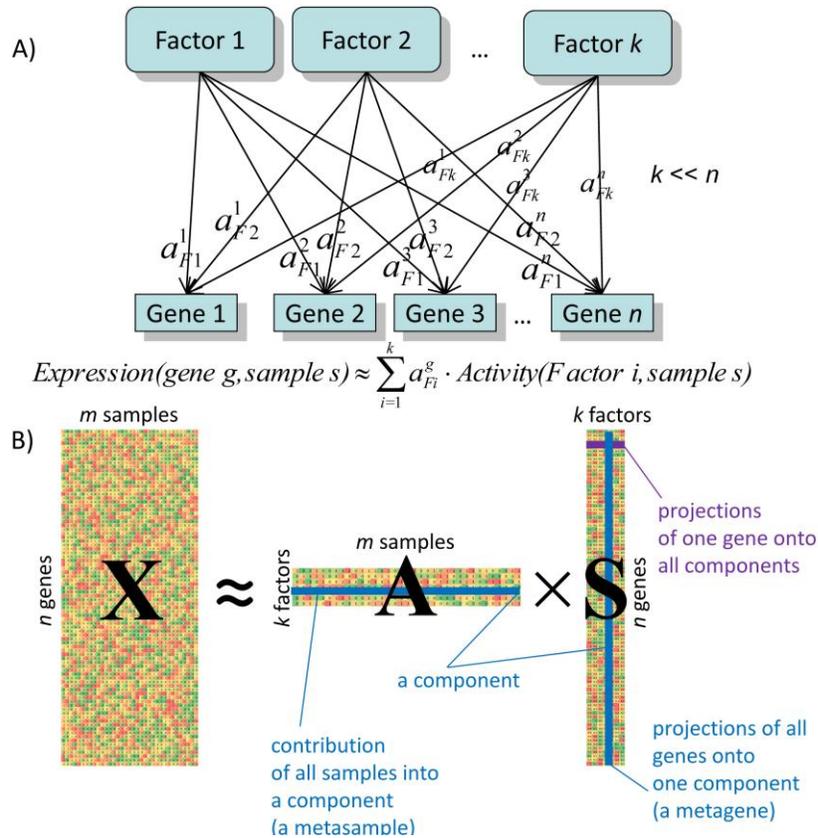

**Figure 2.** Network and matrix representations of linear components. A) Network interpretation of matrix factorization methods. In the "space of genes", each gene expression value is approximated by a weighted linear combination of few factor activities (which can be transcription factors, environmental factors, etc.). Biological samples are characterized by different factor activities, but the weights (network structure) do not depend on samples. In PCA $a^i_j$ implement maximum variance; in ICA $a^i_j$ have maximally non-gaussian distributions for each factor; in NMF $a^i_j \geq 0$. B) A matrix of data $X$ is approximated as a product of low-rank matrices $A$ (mixing matrix) and $S$ (score matrix). Each component corresponds to a row in $A$ (meta-sample) and a column in $S$ (metagene) and introduces two sets of weights for all genes and all samples.

*ICA and NMF outperforms other statistical methods with respect to biological interpretation*

First applications of ICA and NMF demonstrated their outperformance compared to more classical statistical methods (PCA, clustering) with respect to biological interpretation of extracted signals [15]. Similar conclusions were made in [16] for a dataset of endometrial cancers. In [17], the first meta-study comparing ICA with other methods in several large-scale gene expression datasets was performed, including normal human tissues dataset. It was suggested to use the enrichment significance score of Gene Ontology terms in the components as a measure of the method performance, and ICA or NMF outperformed other unsupervised learning methods (PCA, K-Means) in most of the studies.

*Data preconditioning by ICA and NMF improve performance of tumor classification*

In a number of works ICA, NMF and their modifications were used to improve the performance of supervised tumor classification [10, 18, 19, 20, 21]. The general message of these works is that applying classification methods in the space of selected ICA and NMF factors allows either to better stratify a particular cancer into molecular subtypes, or to make better diagnosis and prognosis.

Methods of blind source separation are well-suited for diminishing the effects of non-biological or non-relevant factors on the analysis of expression data that is a major problem in many high-throughput studies [22]. Matrix factorization techniques were already applied for filtering out those signals in gene expression datasets which could be not directly related to tumorigenesis and can confound the biological signals. Thus, Wang et al suggested using ICA decomposition for improving molecular classification of cancers by correcting for tissue heterogeneity and stromal contamination [23]. Recently, Teschendorff et al used ICA to model and correct for presence of known confounding factors in the analysis of microarray data [25]. It was shown that this approach can adjust data analysis for presence of such confounders as beadchip effect and variations in bisulfite conversion efficiency in DNA methylation chips, and for estrogen receptor status and tumour size in gene expression arrays obtained for breast tumour biopsies.

*Data-driven discovery of reproducible biological factors governing tumorigenesis*

The most interesting use of matrix factorization in cancer biology is to find common and universal factors governing tumorigenesis. One of the early attempts of this was undertaken in [24].

For the particular case of breast cancer, independent components extracted for gene expression profiles of breast cancer cell lines were characterized in [26] for cell lines and in [27] for tumors. In [28], several ICA algorithms were tested on 6 breast cancer datasets and the components were systematically interpreted using pathway and Gene Ontology annotations. The most robustly reproduced components from many independent studies were related to immune response, proliferation, stroma and tumor microenvironment. In other cancers, the components connected to specific contaminations (by muscle fibers, epithelial tissue and others) were frequently found (for example, see [29]).

Engreitz et al performed one of the largest gene expression meta-analysis (9395 arrays from 298 public datasets) using ICA which resulted in detection of 423 fundamental components (modules) [30]. Using them, predictions on the mechanism of the anti-cancer drug parthenolide were made.

*Biomarker discovery*

Each ICA or NMF component ranks the genes accordingly to their projection value. This ranking can be utilized for extracting biomarkers [31]. For example, if a component correlates with a tumor classification (for example, good and bad prognosis classes) then the corresponding prognostic markers can be identified.

Discovery of biomarkers is tightly connected to the problem of feature selection for further application of machine learning methods. The problem consists in selecting relatively small but sufficient number of non-redundant and reproducible features (for example, genes or single-nucleotide polymorphisms) to represent a linear component. In practice, this can lead to selecting from several to few tens or hundreds of features per component.

One particular approach for feature selection in matrix factorization is connected to the "gene shaving" idea [32], which consists in iterative removal of the top-scored genes and re-adjusting the components to the remaining data. This approach was applied to cancer gene expression data with PCA components in [32] and for joint analysis of gene expression and copy number data using ICA components in [33].

ICA components were used to define biomarkers in ovarian cancer in [34], and in a meta-study of four different cancers in [35]. To find more robust and biologically meaningful biomarkers, Najarian et al [36] suggested combining the results of ICA analysis and protein-protein interaction networks in one data analysis pipeline.

*Interpretation of linear components*

One of the major challenges in applying linear matrix factorization methods is interpretability of the extracted components. Clustering methods provide their results in the form of relatively small gene sets or groups of samples that can be inspected for the presence of meaningfull biological signals or over-representation of samples of a particular type (in cancer, this can be samples of a particular grade or stage, or samples differently contaminated with a normal tissue, or samples belonging to a particular batch). By contrast, linear components rather corresponds to weigheted collections of many (usually all) genes and samples. Let us outline here the methods used to interpret them.

In mathematical terms, a linear component can be represented in three ways: 1) geometrically, as a basis vector in "gene" or "sample" space (Figure 1) and a set of projection values of genes or samples onto the component; 2) in matrix terms, as a row of the mixing matrix A and the corresponding column of the signal matrix S (Figure 2B); 3) in terms of a network connecting hidden factors and genes, as a profile of the hidden factor activities in samples and connection weights (Figure 2A).

To connect an extracted component with our knowledge of biology, a weigheted set of genes (sometimes called a metagene [10]) can be interpreted using the following methods. Firstly, a threshold on the weight (absolute or signed value) can be used to convert the metagene into a relatively small set of genes for which all "standard" methods of interpretation (including a visual inspection) can be applied, such as a hypergeometrical test for enrichment with a specific gene set (such as a Gene Ontology [37] or a pathway or a set of genes co-localized in the genome). Secondly, the whole set of weights, without introducing a thershold, can be used in Gene Set Enrichment Analysis (GSEA) [38] or similar methods using Kolmogorov-Smirnov like statistical tests. Thirdly, the gene ranking introduced by their weights in a metagene, can be interpreted using interaction networks in analyses such as finding Optimally Functionally Enriched Network (OFTEN) [39] or similar methods.

A weigheted set of samples (sometimes called a meta-sample [35]) can be interpreted using the following ideas. Firstly, one can look at the weight histogram and, if the histogram shows a multimodal distribution, classify biological samples into groups (clusters) that can be interpreted independently, as in "standard" methods (for example, in the standard survival analysis setting). Secondly, one can test if the weights separate significantly or correlate to a particular biological or technical factor (such as patient's age or tumour grade or a particular tumour type or a particular experimental batch).

*Brief comparison of linear factorization methods: PCA, ICA and NMF*

Methods of linear factorization, mentioned in this review, all follow the same principle of approximating a matrix of data by a simpler and low-dimensional matrix. However, there are specific features that can make them suitable or inappropriate in particular applications [13, 14].

First distinctive feature is uniqueness of components. Since PCA corresponds to a simple and unconstrained quadratic optimisation problem, the whole set of principal components corresponds to the global minimum: hense, principal components are unique. By contrast, independent and NMF components realize local minima of the optimisation function that are multiple. In ICA this is because one solves a non-linear optimisation problem while in NMF this is a result of a constrained optimisation problem. To deal with non-uniqueness of the components and their dependence on initial approximations, bootstraping techniques are applied that consist in performing the component extraction many times (hundreds or thousands) and selecting those which are frequently reproduced.

Second importand distinction concerns ranking the components. In PCA, the components are naurally ranked by the proportion of variance they explain. There is no particular natural order in ICA or NMF components, though they can be ordered by a) their reproducibility (robustness) in bootstraps; b) the value of non-Gaussivity function (ICA); c) mutual information estimate between a component and the data; d) their $L_\infty$ norms. In many applications of ICA and NMF, the ordering is not essential, if the data dimension is first reduced using PCA.

Thirdly, as we have already mentioned, ICA and NMF components does not have to be orthogonal in the initial space of data (Figure 1). As a result, they match more closely local data structures (for example, clusters or tails of the multidimensional data distribution).

**Table 1.** A selection of freely available software, implementing ICA and NMF methods, suitable for application to large gene expression datasets.

| Software | Type and Status | Remark | Web-site |
|---|---|---|---|
| **FastICA for Matlab** | Matlab implementation of fast fixed point algorithm for ICA | The robust of components can be computed by bootstraping via accompanying *Icasso* software | http://www.cis.hut.fi/projects/ica/fastica/ <br><br> http://bsp.teithe.gr/members/downloads/Icasso.html |
| **FastICA for R** | R implementation of fastICA | | http://cran.r-project.org/web/packages/fastICA/fastICA.pdf |
| **bioNMF** | Web-based tool | | http://sbm.postech.ac.kr/pna/startAnalysis/onmf/ |
| **BRB-ArrayTools** | Plug-in for BRB-ArrayTools | Implementation of NMF for gene expression analysis | http://linus.nci.nih.gov/BRB-ArrayTools.html <br><br> ftp://linus.nci.nih.gov/pub/NMF |
| **NMF implementation in R** | | Package 'NMF' was removed from the CRAN repository but is available in archive | http://cran.r-project.org/web/packages/NMF/index.html |
| **ONMF** | Web-tool for NMF | Orthogonal NMF, part of Principal Network Analysis tool | http://sbm.postech.ac.kr/pna/startAnalysis/onmf/ |

Last distinction that we mention here concerns non-negativity of NMF. Accordingly to their network interpretation (Figure 2A), ICA and PCA components correspond to biological factors (such as pathway activities) that can affect gene expression with a positive or a negative sign. In this model, a gene expression is a sum of positive and negative influences from different factors, and it suits the best to describe a deviation from some reference expression values (mean expression or gene expression in the normal tissue). NMF components rather model absolute abudances of transcripts: an active factor can positively contribute to the expression value but can not inhibit it. This makes NMF most suitable for deconvolving complex mixtures of molecules, such as in mass spectrometry data, though empirically NMF is able to extract meaningfull signals from gene expression (probably, corresponding to activating or enabling factors).

**Available implementations**

In the Table 1 we list the implementations of the blind source separation methods which are the most used in the analysis of cancer biology data. In this selection, we have focused on the public availability of the implementation, existence of documentation, and scalability for large datasets containing tens of thousands genes whose expression is measured in hundreds of biological samples.

**Summary**

Blind source separation methods have been applied to the analysis of high-throughput molecular biology data in many studies. It was demonstrated several times that the signals obtained by ICA and NMF are more biologically meaningful and reproducible compared other more classical methods (such as PCA, K-means, hierarchical clustering). Methods of blind source separation have a natural interpretation in terms of functioning of regulatory networks (Figure 2A) and take into account pleiotropy of genes. However, detailed and systematic understanding of the hidden signals revealed by these methods and reproduced independently of the technical platforms and specific data collection methods, is still very limited. Much more efforts and experimental validations need to be invested into the detailed interpretation of these hidden factors, governing gene expression and cell functioning.

**Acknowledgements**

This work is supported by the grant INCA LABEL Cancéropole Ile-de-France, by the grant INVADE from ITMO Cancer (Call Systems Biology 2012) and by the grant "Projet Incitatif et Collaboratif Computational Systems Biology Approach for Cancer" from Institut Curie. AZ is a member of the team "Computational Systems Biology of Cancer", Equipe labellisée par la Ligue Nationale Contre le Cancer.